\documentclass[hyper]{JHEP3} 

\JHEPspecialurl{http://jhep.sissa.it/JOURNAL/JHEP3.tar.gz}

\usepackage{epsfig,multicol}

\newcommand\fverb{\setbox\pippobox=\hbox\bgroup\verb}
\newcommand\fverbdo{\egroup\medskip\noindent%
            \fbox{\unhbox\pippobox}\ }
\newcommand\fverbit{\egroup\item[\fbox{\unhbox\pippobox}]}
\newbox\pippobox

\title{Gauged $U(1)_R$ Symmetries and Fayet-Iliopoulos Terms
in 5D Orbifold Supergravity}

\author{Hiroyuki Abe and Kiwoon Choi\\
        Department of Physics,
        Korea Advanced Institute of Science and Technology,
        Daejeon 305-701, Korea\\
E-mail: \email{abe@hep.kaist.ac.kr},
        \email{kchoi@hep.kaist.ac.kr}}

\received{}
\revised{}
\accepted{}
\preprint{\hepth{0412174}\\ KAIST-TH 2004/21}
\dedicated{}

\abstract{We discuss a gauged $U(1)_R$ supergravity on five-dimensional
(5D) orbifold ($S^1/Z_2$) in which both a $Z_2$-even $U(1)$ gauge
field and the $Z_2$-odd graviphoton take part in the $U(1)_R$ gauging.
Based on the off-shell formulation of 5D supergravity, we analyze
the structure of Fayet-Iliopoulos (FI) terms allowed in such model.
Introducing a $Z_2$-even $U(1)_R$ gauge field accompanies new bulk and
boundary FI terms in addition to the known integrable boundary FI term
which could be present in the absence of any gauged $U(1)_R$ symmetry.
Some physical consequences of these new FI terms are examined.}

\dedicated{}

\begin{document}

\section{Introduction}
Supersymmetry (SUSY) is an attractive candidate for physics beyond the
standard model (SM) as it can stabilize the electroweak scale against
the high energy radiative corrections to the Higgs boson mass.
Furthermore SUSY is a fundamental ingredient of the only known consistent
theory of quantum gravity, i.e., superstring theory. To be phenomenologically
viable, SUSY should be broken by a nonzero vacuum value of $F$ and/or $D$
auxiliary component. This can be achieved by introducing a tadpole term for
the $F$ and/or  $D$-component in the effective Lagrangian, for instance
an O'Raifeartaigh term for $F$-breaking~\cite{O'Raifeartaigh:1975pr} and
a Fayet-Iliopoulos (FI) term for $D$-breaking~\cite{Fayet:1974jb}.
However in locally supersymmetric theory $D$-term SUSY breaking is severely
limited since FI term is not allowed {\it unless} the corresponding
$U(1)$ is either an $R$-symmetry~\cite{Barbieri:1982ac} or a so-called
pseudo-anomalous $U(1)$ symmetry with Green-Schwarz anomaly cancellation
mechanism~\cite{Green:1984sg}.

Recently five-dimensional (5D) supergravity (SUGRA) on the orbifold
$S^1/Z_2$  has been studied as an interesting theoretical framework for
physics  beyond the SM. It has been noted that 5D orbifold SUGRA with a
$U(1)_R$ symmetry gauged by the $Z_2$-odd graviphoton can provide the
supersymmetric  Randall-Sundrum (RS) model~\cite{Altendorfer:2000rr}
in which  the weak to Planck scale hierarchy can arise naturally from
the geometric localization of 4D graviton~\cite{Randall:1999ee}.
In this case, the bulk cosmological constant and brane tensions which are
required to generate the necessary AdS$_5$ geometry appear in the Lagrangian
as a consequence of the $U(1)_R$ FI term with $Z_2$-odd coefficient.

In this paper we consider a more generic orbifold  SUGRA which contains
a $Z_2$-even 5D gauge field $A^X_\mu$ participating in the $U(1)_R$ gauging.
If  4D $N=1$ SUSY is preserved by the compactification, the 4D effective
theory of such model will contain a gauged $U(1)_R$ symmetry associated
with the zero mode of $A^X_\mu$, which is not the case when the 5D $U(1)_R$
is gauged only through the $Z_2$-odd graviphoton. Based on the known
off-shell formulation~\cite{Fujita:2001bd}, we formulate a  gauged $U(1)_R$
SUGRA on $S^1/Z_2$ in which both $A^X_\mu$ and the graviphoton take part
in the $U(1)_R$ gauging and then analyze the structure of FI terms
allowed in such model. As expected, introducing a $Z_2$-even $U(1)_R$ gauge
field accompanies new bulk and  boundary FI terms in addition to the known
integrable boundary FI term which could be present in the absence of any gauged
$U(1)_R$ symmetry~\cite{Ghilencea:2001bw,Barbieri:2002ic,
GrootNibbelink:2002wv,Correia:2004pz}.
As we will see, those new FI terms can have interesting implications to the
quasi-localization of the matter zero modes in extra
dimension~\cite{Arkani-Hamed:1999dc}
and also  to the SUSY breaking and radion stabilization.

The organization of this paper is as follows. In the next section
we provide a  formulation for  $U(1)_R$-gauged orbifold SUGRA based on
the Kugo-Ohashi off-shell SUGRA formulation ~\cite{Fujita:2001bd}.
In section~\ref{sec:killing}, we discuss the conditions for unbroken
$N=1$ SUSY for generic 4D Poincar\'e invariant field configurations
and also some physical consequences of the  FI terms of the
$Z_2$-even $U(1)_R$ gauge symmetry.
Section~\ref{sec:conclusion} is a conclusion.

\section{5D orbifold supergravity with gauged $U(1)_R$}
\label{sec:intfi}

In this section we provide a formulation for 5D SUGRA
on $S^1/Z_2$ in which both a $Z_2$-even gauge field  and the $Z_2$-odd
graviphoton take part in the $U(1)_R$ gauging.
As a minimal example, we consider a model containing three $U(1)$
vector multiplets:
\begin{eqnarray}
\begin{array}{rcl}
{\cal V}_Z &=&
\big( \alpha,\, A^Z_\mu,\, \Omega^{Zi},\, Y^{Zij} \big)\,, \\
{\cal V}_X &=&
\big( \beta,\, A^X_\mu,\, \Omega^{Xi},\, Y^{Xij} \big)\,, \\
{\cal V}_S &=&
\big( \gamma, \, A^S_\mu, \,\Omega^{Si},\, Y^{Sij} \big)\,,
\end{array}
\nonumber
\end{eqnarray}
together with a compensator hypermultiplet ${\cal H}_c$
and  physical hypermultiplets ${\cal H}_p$:
\begin{eqnarray}
{\cal H}_c &=&({\cal A}^x_i,\eta^x,{{\cal F}}^x_i)\,,
\nonumber \\
{\cal H}_p &=& (\Phi^x_i,\zeta^x, F^x_i)\,,
\nonumber
\end{eqnarray}
where $M^A=(\alpha,\beta,\gamma)$ ($A=Z,X,S$) are real scalar components,
$\Omega^i_A$ ($i=1,2$) are $SU(2)_{\mathbf{U}}$-doublet symplectic Majorana
spinors, $Y^{Aij}=Y^{Aji}$ are $SU(2)_\mathbf{U}$-triplet auxiliary components,
${\cal A}^x_i,\Phi^x_i$ ($x=1,2$) are quaternionic hyperscalars,
$\eta^x,\zeta^x$ are symplectic Majorana hyperinos, and ${\cal F}^x_i,{F}^x_i$
are the auxiliary components of hypermultiplets.
Here ${\cal V}_Z$ is the model-independent central charge
vector multiplet which contains the $Z_2$-odd graviphoton $A^Z_\mu$,
${\cal V}_X$ is a vector multiplet which contains
a $Z_2$-even $U(1)$ gauge field $A^X_\mu$, and
${\cal V}_S$ is a non-physical vector multiplet which is introduced
to obtain the necessary $Z_2$-odd couplings through the 4-form multiplier
mechanism of Ref.~\cite{Bergshoeff:2000zn}.
The $Z_2$-parities of these components are
summarized in Table~\ref{tab:z2parity}.

Because we have a single compensator hypermultiplet ${\cal H}_c$,
the  corresponding quaternionic manifold spanned by physical hyperscalars
${\cal H}_p$ ($p=1,2,.., n_H$)
is $USp(2,2n_H)/USp(2) \times USp(2n_H)$.
In the following, we will use frequently a matrix
notation for hyperscalars, e.g.,
\begin{eqnarray}
\Phi &\equiv &
\left( \begin{array}{cc}
\Phi^{x=1}_{\ i=1} &
\Phi^{x=1}_{\ i=2} \\
\Phi^{x=2}_{\ i=1} &
\Phi^{x=2}_{\ i=2}
\end{array} \right) \ = \
\left( \begin{array}{cc}
\Phi_+ & \Phi_- \\
-\Phi_-^* & \Phi_+^*
\end{array} \right)\,,
\nonumber
\end{eqnarray}
where $\Phi_{\pm}$ are $Z_2$ parity eigenstates,
and the similar notation is adopted for the
compensator ${\cal A}$. In this matrix notation,
the symplectic reality condition and the $Z_2$
boundary condition are given by
\begin{eqnarray}
\Phi^*(y)=i\sigma_2\Phi(y)i\sigma_2^T\,,
\quad
\Phi(-y)=\sigma_3\Phi (y)\sigma_3\,.
\label{hyperboundary}
\end{eqnarray}
We will also use a matrix notation for the $SU(2)_{\mathbf{U}}$-triplet
fields as, e.g., $Y^A \equiv Y^{Ai}_{\ \ \ j}$.
Note that the $SU(2)_{\mathbf U}$ index is raised and lowered
by an antisymmetric tensor $\epsilon_{ij}=\epsilon^{ij}=i\sigma_2$
like, e.g., $Y^{Ai}_{\ \ \ j}=Y^{Aik}\epsilon_{kj}=\epsilon^{ik}Y^A_{kj}$.

\begin{table}[t]
\begin{center}
\begin{tabular}{c|c|c}
$Z_2$-even & $\alpha$,
$A^{X,S}_{\underline\mu}$, $A^Z_y$,
$(Y^{X,S})^{(3)}$, $(Y^Z)^{(1,2)}$ &
$({\cal A},\Phi)^{x=1}_{i=1}$, $({\cal A},\Phi)^{x=2}_{i=2}$,
$({\cal F},F)^{x=1}_{i=2}$, $({\cal F},F)^{x=2}_{i=1}$ \\ \hline
$Z_2$-odd & $\beta$, $\gamma$,
$A^Z_{\underline\mu}$, $A^{X,S}_y$,
$(Y^Z)^{(3)}$, $(Y^{X,S})^{(1,2)}$ &
$({\cal A},\Phi)^{x=1}_{i=2}$, $({\cal A},\Phi)^{x=2}_{i=1}$,
$({\cal F},F)^{x=1}_{i=1}$, $({\cal F},F)^{x=2}_{i=2}$ \\
\end{tabular}
\end{center}
\caption{The $Z_2$-parity assignment of component fields.
Here $\underline\mu=0,1,2,3$, while  $y$ is the fifth coordinate of
$S^1/Z_2$ and $Y^{Ai}_{\ \ \ j}=\sum_{r=1}^3 (i\sigma_r)^i_{\ j}
(Y^A)^{(r)}$ $(A=Z,X,S)$.}
\label{tab:z2parity}
\end{table}

The most general hypermultiplet gauging consistent with
the $Z_2$ orbifolding and the symplectic reality condition
(\ref{tab:z2parity}) is given by
\begin{eqnarray}
\Big(\,T_Z,T_X,T_S\,\Big)\Phi&=&
\Big( 0,\,q,\,c \Big)i\sigma_3\Phi\,,
\nonumber \\
\Big(\,T_Z,T_X,T_S\,\Big){\cal A}&=&
\Big( 0,\,-r,\, \textstyle{-\frac{3}{2}}k \Big)i\sigma_3{\cal A}\,,
\nonumber
\end{eqnarray}
where $T_{Z,X,S}$ are the $U(1)_{Z,X,S}$ generators
and $q,c,r,k$ are real constants.
The vector multiplet part of 5D SUGRA is determined
by the norm function ${\cal N}$ which is a homogeneous cubic polynomial
of $M^A$:
${\cal N}=C_{ABC}M^AM^BM^C$ for a totally symmetric constants $C_{ABC}$.
In this paper, we choose
\begin{eqnarray}
{\cal N} &=& \alpha^3 -\frac{1}{2} \alpha \beta^2
+\frac{1}{2} \xi_{FI} \alpha \beta \gamma\,,
\nonumber
\end{eqnarray}
which corresponds to a minimal model for our purpose.

Under the constraint on ${\cal V}_S$ induced by the four-form multiplier
field which will be introduced below, a nonzero $\xi_{FI}$ gives an integrable 
boundary FI term for the $U(1)_X$ vector multiplet:
$-\frac{1}{2}\xi_{FI}Y^{X\,(3)}\partial_y\epsilon(y)$ where we use an
isovector notation $Y^A=\sum_{r=1}^3(i\sigma_r)Y^{A\,(r)}$,
the gauge charge $k$ of the compensator  gives
a negative bulk cosmological constant $-6k^2$ for AdS$_5$ geometry as
well as the correct integrable boundary tension $3k\partial_y\epsilon(y)$,
and finally the gauge charge $c$ of the physical hypermultiplet
gives a hyperino kink mass $c\epsilon(y)$.
The four-form multiplier mechanism of Ref.~\cite{Bergshoeff:2000zn} provides a
dynamical way to generate $Z_2$-odd couplings proportional to the periodic
sign-function $\epsilon(y)=y/|y|$ which obeys
\begin{eqnarray}
&& \epsilon(y)=-\epsilon(-y)=\epsilon(y+2\pi R)=1 \quad (0<y<\pi R)\,,
\nonumber \\
&& \partial_y\epsilon(y)=2\left[\,\delta(y)-\delta(y-\pi R)\,\right]\,.
\nonumber
\end{eqnarray}
To implement the four-form mechanism within off-shell SUGRA,
we introduce a multiplier multiplet defined under a background of ${\cal V}_Z$:
\begin{eqnarray}
L_H &=&
(L^{ij},\,\varphi^i,\,E_{\mu \nu \rho},\,
H_{\mu \nu \rho \sigma})\,,
\label{multi}
\end{eqnarray}
where $L^{ij}$ is a $SU(2)$-triplet scalar,
$\varphi^i$ is a $SU(2)$-doublet fermion, $E_{\mu \nu \rho}$
and $H_{\mu \nu \rho \sigma}$ are three and four-form fields, respectively.
The off-shell action of this multiplier multiplet was derived
in Ref.~\cite{Fujita:2001bd} as
\begin{eqnarray}
{\cal L}^{\rm bulk}_{\rm 4\textrm{-}form} &=&
e(Y^{Sij}-GY^{Zij})L_{ij}
-\frac{1}{4!}\epsilon^{\lambda \mu \nu \rho \sigma} \left[ \left\{
F_{\lambda \mu}(A^S) -GF_{\lambda \mu}(A^Z) \right\} E_{\nu \rho \sigma}
+\frac{1}{2} G \partial_\lambda H_{\mu \nu \rho \sigma} \right]\,,
\nonumber \\
{\cal L}^{\rm brane}_{\rm 4\textrm{-}form} &=&
\left( a_0 \delta(y)+a_\pi\delta(y-\pi R) \right)
\left[ \frac{1}{4!}
\epsilon^{\mu \nu \rho \sigma y} H_{\mu \nu \rho \sigma}
+e_{(4)}\alpha \epsilon^{jk}(i\sigma_3)^i_{\ k}L_{ij} \right]\,,
\label{eq:bkvpbrane}
\end{eqnarray}
where $\epsilon^{jk}$ is the $SU(2)$-invariant antisymmetric tensor,
$G=M^S/M^Z=\gamma/\alpha$, $e=(-
\mbox{det}(g_{\mu\nu}))^{1/2}$, and
$e_{(4)}=(-\mbox{det}(g_{\underline\mu\underline\nu}))^{1/2}$
for the induced 4D metric $g_{\underline\mu\underline\nu}$ on the boundaries.
Then the equations of motion for $H_{\mu \nu \rho \sigma}$,
$E_{\mu \nu \rho}$ and $L_{ij}$ give
\begin{eqnarray}
G \ &=& \ \gamma/\alpha = \epsilon(y)\,,
\nonumber\\
F_{\mu \nu}(A^S) &=& \epsilon(y)F_{\mu \nu}(A^Z)\,,
\nonumber \\
Y^{Sij} &=& \epsilon(y)Y^{Zij}
+e^{-1}e_{(4)} \alpha \epsilon^{jk}(i\sigma_3)^i_{\ k}
(\delta(y)-\delta(y-\pi R))\,,
\label{eq:yrel}
\end{eqnarray}
where the integrability condition of $\partial_y G$ results in
$a_0=-a_\pi = -1/2$.
Now using the relations in (\ref{eq:yrel}), the redundant
vector multiplet ${\cal V}_S$ can be {\it replaced} by the central charge
vector multiplet ${\cal V}_Z$ multiplied by the $Z_2$-odd factor $\epsilon(y)$.
This four-form mechanism provides an elegant way to obtain a locally
supersymmetric theory of ${\cal V}_I$ ($I=Z,X$) involving $Z_2$-odd couplings,
e.g., $\xi_{FI}\epsilon(y), k\epsilon(y)$ and $c\epsilon(y)$ in our case,
starting from a locally supersymmetric theory of ${\cal V}_A$ ($A=Z,X,S$) and
the four-form multiplier multiplet involving only $Z_2$-even couplings.

Upon ignoring the UV-sensitive higher-dimensional boundary terms,
after integrating out the Lagrange multipliers and also the auxiliary fields
{\it other than} $Y^{Iij}$, $V^{ij}_\mu$,
$v_{\mu\nu}$ and ${\cal F}^x_i$ (here $V^{ij}_\mu$ and $v_{\mu\nu}=-v_{\nu\mu}$
are the auxiliary fields in the Weyl multiplet of 5D conformal
SUGRA),
we find the following Lagrangian density of bosonic fields:
\begin{eqnarray}
{\cal L} &=&
{\cal L}_{\rm bulk} + {\cal L}_{\rm brane} + {\cal L}_{\rm aux},
\nonumber \\
e^{-1}{\cal L}_{\rm bulk} &=&
-\frac{1}{2}R
-\frac{1}{4}\tilde{a}_{IJ}
F^I_{\mu \nu} F^{\mu \nu J}
+\frac{1}{2}\tilde{a}_{IJ}
\nabla_\mu M^I \nabla^\mu M^J
\nonumber \\
&&+
\frac{1}{8}e^{-1} \tilde{C}_{IJK}
\epsilon^{\lambda \mu \nu \rho \sigma} A^I_\lambda
 F^J_{\mu \nu} F^K_{\rho \sigma}
+{\rm tr} \Big[ \nabla_\mu \Phi\nabla^\mu\Phi^{\dagger}-
\nabla_\mu {\cal A}\nabla^\mu {\cal A}^{\dagger}
\nonumber \\
&&-V^\dagger_{\mu\,{\rm sol}}V^\mu_{\rm sol}
-M^IM^J \big(
\Phi^\dagger {t}_I^\dagger {t}_J \Phi
-{\cal A}^\dagger {t}_I^\dagger {t}_J {\cal A}
\big) \Big]
\nonumber \\ &&
-\frac{1}{2} {\rm tr}\Big[
\tilde{\cal N}_{IJ} Y^{I\dagger}Y^J
-4Y^{I\dagger}
\left( {\cal A}^\dagger {t}_I {\cal A}
-\Phi^\dagger {t}_I \Phi \right)\Big] \,,
\label{lbulk} \\
e_{(4)}^{-1} {\cal L}_{\rm brane} &=&  \Bigg[ \,
\frac{1}{2}\xi_{FI}\alpha^2
\Big(\, {\rm tr}[\,i\sigma_3 Y^X\,] +e^{-1}e_{(4)}\partial_y\beta \, \Big)
+\frac{1}{2}\xi_{FI}\alpha \beta\, {\rm tr}[\,i\sigma_3 Y^Z\,]
\nonumber \\ && \quad
-2\alpha\left(3k+\frac{3}{2}k\, {\rm tr} \left[ \Phi^\dagger \Phi \right]
+c\, {\rm tr} \left[ \Phi^\dagger \sigma_3 \Phi
\sigma_3 \right] \right) \Bigg]
\left( \delta(y)-\delta(y-\pi R) \right)\,.
\label{eq:lbrane} \\
e^{-1}{\cal L}_{\rm aux} &=&
-(V_\mu-V_{\mu\,{\rm sol}})^{ij}(V^\mu-V^\mu_{\rm sol})_{ij}
+(1-\alpha^{-2} A^Z_\mu A^{Z\mu})
\rm{tr}\left[({\cal F}-{\cal F}_{\rm sol})^\dagger
({\cal F}-{\cal F}_{\rm sol})\right]
\nonumber \\ &&
+2(v-v_{\rm sol})^{\mu\nu}(v-v_{\rm sol})_{\mu\nu}\,,
\label{eq:laux}
\end{eqnarray}
where
\begin{eqnarray}
\tilde{\cal N} &=&
\left.{\cal N}\right|_{\gamma=\epsilon(y)\alpha} \ = \
\tilde{C}_{IJK}M^IM^JM^K
=\alpha^3 -\frac{1}{2} \alpha \beta^2
+\frac{1}{2}\xi_{FI} \epsilon(y) \alpha^2 \beta,
\nonumber \\
\tilde{a}_{{I}{J}}
&=& -\frac{1}{2} \frac{\partial^2\ln\tilde{\cal N}}
{\partial M^I\partial M^J}
\ = \ \left( \begin{array}{cc}
\tilde{a}_{ZZ} & \tilde{a}_{ZX} \\
\tilde{a}_{XZ} & \tilde{a}_{XX}
\end{array} \right),
\nonumber \\
\tilde{a}_{ZZ} &=&
\frac{1}{8\tilde{\cal N}^2}\big(
\beta^4 +2\alpha (6\alpha^3
+4\xi_{FI}\epsilon(y)\alpha^2\beta
-\xi_{FI}\epsilon(y)\beta^3
+\xi_{FI}^2\epsilon^2(y)\alpha \beta^2) \big),
\nonumber \\
\tilde{a}_{XZ} &=& \tilde{a}_{ZX} \ = \
\frac{1}{8\tilde{\cal N}^2}\alpha^2 \big(
-8\alpha\beta+\xi_{FI}\epsilon(y)(
2\alpha^2-\beta^2) \big),
\nonumber \\
\tilde{a}_{XX} &=&
\frac{1}{8\tilde{\cal N}^2}\alpha^2 \big(
2\beta^2-2\xi_{FI}\epsilon(y)\alpha\beta
+(4+\xi_{FI}^2\epsilon^2(y))\alpha^2 \big),
\nonumber \\
V_{\mu\,{\rm sol}} &=&
-\frac{1}{2}\left( {\cal A}^\dagger (\nabla_\mu {\cal A})
- (\nabla_\mu {\cal A})^\dagger {\cal A} \right)
+\frac{1}{2}\left( \Phi^\dagger (\nabla_\mu \Phi)
- (\nabla_\mu \Phi)^\dagger \Phi \right),
\label{eq:osvy} \\
{\cal F}_{\rm sol} &=& \alpha t_Z {\cal A}, \qquad
v_{\mu\nu\,{\rm sol}} \ = \ -\frac{1}{4\tilde{\cal N}}
\tilde{\cal N}_IF_{\mu\nu}(A^I)\,.
\nonumber
\end{eqnarray}
Here $\tilde{\cal N}_I=\frac{\partial{\tilde{\cal N}}}{\partial M^I}$,
$\tilde{\cal N}_{IJ}=\frac{\partial^2{\tilde{\cal N}}}{\partial M^I\partial M^J}$
for $M^I=(\alpha,\beta)$ ($I=Z,X$) and
$$
\tilde{\cal N}^{{I}{J}}
=\frac{1}{\beta^2-\xi_{FI}\epsilon(y)\alpha\beta
+\big(6+\xi_{FI}^2\epsilon^2(y)\big)\alpha^2}
\left( \begin{array}{cc}
\alpha & -\beta+\xi_{FI}\epsilon(y)\alpha \\
-\beta+\xi_{FI}\epsilon(y)\alpha &
-6\alpha-\xi_{FI}\epsilon(y)\beta
\end{array} \right)\,,
$$
is the inverse matrix of $\tilde{\cal N}_{IJ}$.
Note that after the multiplier multiplet (\ref{multi}) is integrated out,
the new $U(1)$ generators $t_I$ for hyperscalars are given by
\begin{eqnarray}
\Big( {t}_Z,\,{t}_X \Big)\Phi &=&
\Big( c\epsilon(y),\, q \Big)i\sigma_3\Phi\,,
\nonumber \\
\Big( {t}_Z,\,{t}_X \Big){\cal A}&=&
\Big( \textstyle{-\frac{3}{2} k \epsilon(y)},\, -r \Big)i\sigma_3{\cal A}\,,
\label{eq:u1charge}
\end{eqnarray}
Here we have already
integrated out the auxiliary fields other than $Y^{Iij}$, $V^{ij}_\mu$,
$v_{\mu\nu}$ and ${\cal F}^x_i$
which we keep at off-shell values
since the on-shell values of these auxiliary fields are
affected by the boundary supergravity which will be discussed at the end
of this section. The complete form of ${\cal L}_{\rm aux}$ involving all
auxiliary components can be found in Ref.~\cite{Fujita:2001bd}.

The $2\times 2$ matrix valued compensator hyperscalar field can be chosen as
\begin{eqnarray}
{\cal A} &\equiv&  \mathbf{1}_2\,
\sqrt{1+\frac{1}{2}{\rm tr} [\Phi^\dagger \Phi]}\,,
\label{eq:s:ufix}
\end{eqnarray}
in the unit with the 5D Planck mass $M_5=1$,
which corresponds to the $SU(2)_{\mathbf{U}}$ gauge fixing condition
in the hypermultiplet compensator formulation of off-shell
5D SUGRA~\cite{Fujita:2001bd}. Also by the dilatation gauge fixing
condition the norm function can be fixed as
\begin{eqnarray}
\tilde{{\cal N}} &\equiv& 1\,.
\nonumber
\end{eqnarray}
 Then  we have only
one physical gauge scalar field $\phi$ in our system which parameterizes
the original scalar fields $\alpha$ and $\beta$ in $\tilde{\cal N}$ as
\begin{eqnarray}
\alpha &=&
\frac{\cosh^{2/3} (\phi)}{(1+\xi_{FI}^2\epsilon^2(y)/8)^{1/3}}
\nonumber \\
&=&1+\frac{1}{3}\Big(
\phi^2-\frac{1}{8}\xi_{FI}^2\epsilon^2(y)\Big)
+{\cal O}(\phi^4)\,,
\nonumber \\
\beta &=&
\frac{\cosh^{2/3} (\phi)[ (2+\xi_{FI}^2\epsilon^2(y)/4)^{1/2}
\tanh (\phi) + \xi_{FI} \epsilon(y)/2 ]}{(
1+\xi_{FI}^2\epsilon^2(y)/8)^{1/3}}
\nonumber \\
&=& \frac{1}{2}\xi_{FI}\epsilon(y)+\sqrt{2}\phi+
{\cal O}(\phi^3)\,.
\nonumber
\end{eqnarray}
The very special manifold spanned by $\phi$ has the metric
\begin{eqnarray}
g_{\phi\phi}(\phi) &=& -\frac{1}{2} \frac{\partial^2\ln\tilde{\cal N}}
{\partial M^I\partial M^J}
\frac{\partial M^I}{\partial \phi}
\frac{\partial M^J}{\partial \phi}
\nonumber \\
&=&
\frac{1+2\cosh(2\phi)}{3\cosh^2(\phi)}
\,=\,1+\frac{1}{3}\phi^2+{\cal O}(\phi^4)\,.
\nonumber
\end{eqnarray}
Obviously, $\phi$ is $Z_2$-odd for $Z_2$-odd $\beta$.

Let us identify the gauged $U(1)_R$ symmetries of the model.
We note that $SU(2)_R$ representation is labelled by the $i,j$ indices of
component fields {\it after} the $SU(2)_{\mathbf{U}}$ gauge fixing
(\ref{eq:s:ufix}). In this setting, gauging  the $R$-symmetry corresponds
to making the compensator hypermultiplet to have a nonzero gauge coupling.
If the compensator couples to a physical gauge field $A^R_\mu$, the covariant
derivative is given by
\begin{eqnarray}
{\cal D}_\mu {\cal A}^x_{\ i} &=& \partial_\mu {\cal A}^x_{\ i}
-(V_\mu)_{ij} {\cal A}^{xj} - (A^R_\mu)^x_{\ y} {\cal A}^y_{\ i}\,,
\nonumber
\end{eqnarray}
where $(A^R_\mu)^x_{\ y}=-(rA^X_\mu+\frac{3}{2}k\epsilon(y)A^Z_\mu)
(i\sigma_3)^x_{\ y}$ in our case, and $V_\mu$ is the auxiliary
$SU(2)_{\mathbf{U}}$  gauge field in the SUGRA (Weyl) multiplet.
After the compensator gauge fixing (\ref{eq:s:ufix}), the auxiliary
$SU(2)_{\mathbf{U}}$  gauge field is redefined as~\cite{Kugo:2000af}
\begin{eqnarray}
(V^N_\mu)^i_{\ j} &=& (V_\mu)^i_{\ j}-(A^R_\mu)^i_{\ j}=
(V_\mu)^i_{\ j}
+\left(rA^X_\mu+\frac{3}{2}k\epsilon(y)A^Z_\mu\right)(i\sigma_3)^i_{\ j}\,,
\label{redef:aux}
\end{eqnarray}
yielding the $R$-gauge couplings of all
$SU(2)_\mathbf{U}$ non-singlet physical fields,
e.g.,
\begin{eqnarray}
{\cal D}_\nu \psi^i_\mu &=&
\nabla_\nu \psi^i_\mu -(V_\nu)^i_{\ j} \psi^j_\mu
\nonumber \\ &=&
\nabla_\nu \psi^i_\mu -(V^N_\nu)^i_{\ j} \psi^j_\mu
+\left(rA^X_\nu+\frac{3}{2}k\epsilon(y)A^Z_\nu\right)
(i\sigma_3)^i_{\ j} \psi^j_\mu \,,
\nonumber
\end{eqnarray}
for the gravitino $\psi^i_\mu$.
Therefore when $r\neq 0$,  the $Z_2$-even $A^X_\mu$ becomes
a $U(1)_R$ gauge field in the $\sigma_3$ direction of $SU(2)_R$, while for
$k\neq 0$ the $Z_2$-odd graviphoton $A^Z_\mu$ becomes a $U(1)_R$ gauge field
again in the $\sigma_3$ direction.
In this prescription, the $Z_2$-even (odd) hyperscalar $\Phi_+$ ($\Phi_-$)
carries a $U(1)_X$ charge $q+r$ ($q-r$), while its fermionic partner carries
a $U(1)_X$ charge $q$.

The model described by (\ref{lbulk}), (\ref{eq:lbrane}) and (\ref{eq:laux})
contain various FI terms which are linear in the auxiliary components
$Y^{Z,X}$. For instance, there appear the boundary FI terms of $Y^{X,Z}$
which arise from the $\xi_{FI}$-term in $\tilde{\cal N}$~\cite{Barbieri:2002ic}:
\begin{eqnarray}
\frac{1}{2}\xi_{FI}\left[
\alpha^2\left( {\rm tr}[i\sigma_3Y^{X}]+e^y_4 \partial_y\beta \right)
+\alpha\beta{\rm tr}[i\sigma_3 Y^Z]\right]
\big( \delta(y)-\delta(y-\pi R) \big)\,,
\label{eq:ifi}
\end{eqnarray}
The first $U(1)_X$ FI term has been discussed
extensively in the literatures~\cite{Ghilencea:2001bw,Barbieri:2002ic,
GrootNibbelink:2002wv,Correia:2004pz}
together with its physical consequences. (In Ref.~\cite{Correia:2004pz}
this FI term is derived in a simpler way by utilizing a superfield
approach~\cite{PaccettiCorreia:2004ri} to 5D conformal supergravity.)
As for the second FI term of $Y^Z$, it involves the product of two
$Z_2$-odd fields, $\beta$ and $Y^{Z\,(3)}$,
and thus depends on how to regulate the behavior of these $Z_2$-odd
fields across the boundary. However, while the first term is
${\cal O}(\xi_{FI})$, the second term is  ${\cal O}(\xi_{FI}^2)$ since
$\beta={\cal O}(\xi_{FI})$ near the boundary, thus can be ignored when
$\xi_{FI}\ll M_5$ which is the limit that orbifold SUGRA does make sense.

In case with  gauged  $U(1)_R$ symmetry, there exist additional FI terms
as expected.
When $r\neq 0$ and/or $k\neq 0$,
the term $2{\rm tr} \big[ Y^{I\dagger}{\cal A}^\dagger t_I{\cal A} \big]$
in our bulk Lagrangian (\ref{lbulk}) gives a bulk FI term for $Y^X$ and/or
$Y^Z$ after the gauge fixing (\ref{eq:s:ufix}):
\begin{eqnarray}
2r\,{\rm tr}\big[i\sigma_3 Y^X \big]+3k\epsilon(y){\rm tr}\big[
i\sigma_3 Y^Z\big]\,.
\label{eq:bfi}
\end{eqnarray}
Integrating out $Y^Z$ then leads to a bulk cosmological constant
$-3k^2/2\alpha$ with which the total bulk cosmological constant
is given by $-k^2(3\alpha^{-1}+9\alpha^2)/2=-6k^2+{\cal O}(k^2\phi^2)$.
Also as a consequence of (\ref{eq:yrel}), the bulk FI term of $Y^Z$ appears
together with the boundary tension term:
$3\alpha k\partial_y\epsilon(y)=(3k+{\cal O}(k\phi^2))\partial_y\epsilon(y)$.
When $r\neq 0$, there can be additional boundary FI terms of $Y^X$.
To see this, let us briefly review the construction of 4D SUGRA
at the boundaries in 5D orbifold SUGRA~\cite{Fujita:2001bd}.
The general boundary Lagrangian can be written as
\begin{eqnarray}
{\cal L}_{\rm N=1} = \sum_{l=0,\pi}
\Lambda_l \delta(y-y_l)
\left( {\textstyle -\frac{3}{2}}
\big[ \Sigma \bar\Sigma e^{-K^{(l)}(S,\bar{S})/3} \big]_D
+ \big[ f^{(l)}_{IJ} (S) W^{I \alpha} W^J_\alpha \big]_F
+ \big[ \Sigma^3 W^{(l)}(S) \big]_F \right)\,,
\label{boundarysugra}
\end{eqnarray}
where $(y_0,y_\pi)=(0,\pi R)$ and $\Lambda_{0,\pi}$ are constants.
$\Sigma$ is the 4D $N=1$ compensator chiral multiplet induced
by the 5D compensator hypermultiplet ${\cal H}_c$,
and $S$ and $W^{I\alpha}$ stand for generic chiral matter and gauge
multiplets at the boundaries which come from either bulk fields or
pure boundary fields. Here the subscripts $D$ and $F$ represent
the $D$- and $F$-components, respectively, in the
4D superconformal tensor calculus~\cite{Fujita:2001bd}.
Let $\Sigma=( z^0, \chi^0_R, F^0 )$ denote the component fields of the
$N=1$ compensator  superfield.
The bosonic components of $\Sigma$ are given by~\cite{Fujita:2001bd}
\begin{eqnarray}
z^0 &=&\big( {\cal A}_+^* \big)^{2/3},
\nonumber \\
F^0 &=& \frac{2}{3}i({\cal A}_+^*)^{2/3}
e^y_4\left( V_y^{N\,(1)}+iV_y^{N\,(2)} \right)
+\frac{2}{3} \left( 1+i\alpha^{-1}e^y_4A^Z_y \right)({\cal A}_+^*)^{-1/3}
\left( \tilde{\cal F}^{(1)}+i\tilde{\cal F}^{(2)} \right)\,,
\nonumber
\end{eqnarray}
where ${\cal A}_+\equiv ({\cal A}^{x=2}_{i=2})^*
=\sqrt{1+{\rm tr}[\Phi^\dagger \Phi]/2}$,
$V^N_\mu$ is the auxiliary $SU(2)_{\bf U}$ gauge field redefined as
(\ref{redef:aux}), and
$\tilde{\cal F}\equiv {\cal F}-{\cal F}_{\rm sol}
={\cal F}-\alpha t_Z{\cal A}$.
Here we use the notation
$V^N_\mu=\sum_{r=1}^3(i\sigma_r)V_\mu^{N\,(r)}$ and
$\tilde{\cal F}=\tilde{\cal F}^{(0)}\mathbf{1}_2
-\sum_{r=1}^3(i\sigma_r)\tilde{\cal F}^{(r)}$.
Using the standard tensor calculus, it is straightforward to find
the $D$-component of the real superfield $\Sigma\bar\Sigma$:
\begin{eqnarray}
\left. \Sigma\bar\Sigma \right|^{\rm bosonic}_D
=\frac{4}{3}r|z^0|^2(2Y^{X\,(3)}-e^y_4 \partial_y\beta)
+2 \left( |F^0|^2
+|\hat{\cal D}^{(4)}_{\underline{\mu}} z^0|^2 \right)\,,
\nonumber
\end{eqnarray}
where
$$
\hat{\cal D}^{(4)}_{\underline{\mu}} z^0
=\frac{2}{3}z_0^{-1/2}
\Big( \partial_{\underline{\mu}}{\cal A}_+^*
-i(V_{\underline{\mu}}^{N\,(3)}+e^y_4v_{\underline{\mu}y})
{\cal A}_+^* \Big)\,.
$$
Here we omit all the fermionic degrees of freedom
and also ignored the terms involving $Z_2$-odd fields which
either vanish at the boundaries or are irrelevant for the following discussion.

As a minimal example of the boundary SUGRA, let us consider the pure 4D SUGRA
on the boundaries which corresponds to the case with
$K^{(0,\pi)}=f^{(0,\pi)}_{IJ}=W^{(0,\pi)}=0$
in (\ref{boundarysugra}).
Then the bosonic part of the boundary SUGRA is given by
\begin{eqnarray}
{\cal L}_{N=1} &=&
-\frac{3}{2}
\left[ \Sigma \bar\Sigma \right]_D^{\rm bosonic}
\big( \Lambda_0 \delta(y) + \Lambda_\pi \delta(y-\pi R) \big)
\nonumber \\ &=&
-e_{(4)} M_{(4)}^2 \Bigg[\,\frac{1}{2} R^{(4)}+
2r(2Y^{X(3)}-e_4^y \partial_y \beta)
\nonumber \\ &&
+\,3 |z^0|^{-2} \left( |F^0|^2
+|\hat{\cal D}^{(4)}_{\underline\mu} z^0|^2 \right)
\Bigg] \big( \Lambda_0 \delta(y) + \Lambda_\pi \delta(y-\pi R) \big)\,,
\label{minimal}
\end{eqnarray}
where $R^{(4)}$ is the induced Ricci-scalar on the boundaries and
\begin{eqnarray}
M_{(4)}^2 &=&
\Big(M_5^3+{\rm tr}[\Phi^\dagger \Phi]/2\Big)^{2/3}\,,
\label{eq:bigs}
\end{eqnarray}
where we recover the 5D Planck mass $M_5$ which was set
as $M_5=1$ in the previous discussion.
The above boundary SUGRA contains a Fayet-Iliopoulos term
\begin{eqnarray}
-2rM_{(4)}^2 \left(2Y^{X\,(3)}-e_4^y \partial_y \beta\right)
\big( \Lambda_0 \delta(y) + \Lambda_\pi \delta(y-\pi R) \big)\,.
\label{eq:nfi}
\end{eqnarray}
Note that unlike the boundary FI term (\ref{eq:ifi}) from the
$\xi_{FI}$-term in $\tilde{N}$,
these boundary FI terms from $U(1)_R$ gauging  can have independent
coefficients at different boundaries, i.e.
$\Lambda_{0}$ and $\Lambda_\pi$ are independent from each other.

The total action of our $U(1)_R$-gauged SUGRA including the minimal
boundary SUGRA (\ref{minimal})
is given by
\begin{eqnarray}
{\cal L} &=&
{\cal L}_{\rm bulk}+{\cal L}_{\rm brane}
+{\cal L}_{\rm aux}+{\cal L}_{N=1}\,.
\label{eq:fullaction}
\end{eqnarray}
In the presence of ${\cal L}_{N=1}$,
the on-shell values of the auxiliary fields $Y^I$, $V^N_\mu$,
$\tilde{\cal F}={\cal F}-\alpha t_Z{\cal A}$, $v_{\mu\nu}$ are found to be
\begin{eqnarray}
Y^I &=&\tilde{\cal N}^{IJ}\tilde{\cal Y}_J,
\quad
V_{\underline{\mu}}^{N\,(1),(2)}
\ = \ V_{\underline{\mu}\,{\rm sol}}^{(1),(2)}, \quad
V_{y}^N \ = \ V_{y\,{\rm sol}}+\Delta V_{y\,{\rm sol}},
\nonumber \\
V_{\underline{\mu}}^{N\,(3)} &=&
\frac{2+\Delta_D}{2+3\Delta_D}
V^{(3)}_{\underline{\mu}\,{\rm sol}}
-\frac{2\Delta_D}{2+3\Delta_D}
e^y_4v_{\underline{\mu}y\,{\rm sol}},
\quad \tilde{\cal F}^{(0),(3)} \ = \ 0,
\nonumber \\
\tilde{\cal F}^{(1)} &=&
\frac{{\cal A}_+^{-1}\Delta_D}
{1+({\cal A}_+^{-2}-1) \Delta_D}
\frac{e_4^y(V_{y\,{\rm sol}}^{(2)}
-\alpha^{-1}e_4^y A^Z_y V_{y\,{\rm sol}}^{(1)})}
{1-\alpha^{-2}A^Z_\mu A^{Z\mu}},
\nonumber \\
\tilde{\cal F}^{(2)} &=&
-\frac{{\cal A}_+^{-1}\Delta_D}
{1+({\cal A}_+^{-2}-1) \Delta_D}
\frac{e^y_4(V_{y\,{\rm sol}}^{(1)}
+\alpha^{-1}e^y_4 A^Z_y V_{y\,{\rm sol}}^{(2)})}
{1-\alpha^{-2}A^Z_\mu A^{Z\mu}},
\nonumber \\
v_{\underline{\mu}\underline{\nu}}
&=& v_{\underline{\mu}\underline{\nu}\,{\rm sol}}, \qquad
v_{\underline{\mu}y} \ = \
\frac{2+2\Delta_D}{2+3\Delta_D}
v_{\underline{\mu}y\,{\rm sol}}
-\frac{\Delta_D}{2+3\Delta_D}
e_y^4V^{(3)}_{\underline{\mu}\,{\rm sol}}\,,
\label{eq:osaux}
\end{eqnarray}
where
\begin{eqnarray}
\tilde{\cal Y}_I &=&
2({\cal A}^\dagger (t_I) {\cal A}-\Phi^\dagger (t_I) \Phi)
-\frac{1}{2}\xi_{FI}e^{-1}e_{(4)}(i\sigma_3)
\alpha^2 \delta_I^{\ X}
\big( \delta(y)-\delta(y-\pi R) \big)
\nonumber \\ &&
-2r M_{(4)}^2 (i\sigma_3) \delta_I^{\ X}
\big( \Lambda_0 \delta(y)+ \Lambda_\pi \delta(y-\pi R) \big),
\nonumber \\
\Delta V_{y\,{\rm sol}} &=&
\frac{\Delta_D}
{1+({\cal A}_+^{-2}-1) \Delta_D}
\Big( (i\sigma_1)V_{y\,{\rm sol}}^{(1)}
+(i\sigma_2)V_{y\,{\rm sol}}^{(2)} \Big),
\nonumber \\
\Delta_D &=& \frac{2}{3}e^{-1}e_{(4)}M_{(4)}^2
\big( \Lambda_0 \delta(y) +\Lambda_\pi \delta(y-\pi R) \big)\,,
\label{eq:dv4sol}
\end{eqnarray}
and $V_{\mu\,{\rm sol}}=\sum_{r=1}^3 (i\sigma_3)V_{\mu\,{\rm sol}}^{(r)}$
and $v_{\mu \nu\,{\rm sol}}$ are defined as  Eq.~(\ref{eq:osvy}).
The part of (\ref{eq:fullaction}) which corresponds to the 5D scalar
potential is given by
\begin{eqnarray}
V_{5D}&=& \,{\rm tr}\Big[
M^IM^J \big\{ \Phi^\dagger {t}_I^\dagger {t}_J \Phi
-{\cal A}^\dagger {t}_I^\dagger {t}_J {\cal A} \big\}
-\frac{1}{2} \tilde{\cal N}^{IJ}
\tilde{\cal Y}_I^\dagger \tilde{\cal Y}_J \Big]
\nonumber \\
&&+\, 2e^{-1}e_{(4)} \alpha \left(
3k+\frac{3}{2} k\,{\rm tr} \left[ \Phi^\dagger \Phi \right]
+c\,{\rm tr} \left[ \Phi^\dagger \sigma_3 \Phi
\sigma_3\right] \right) \left(
\delta (y)-\delta(y-\pi R)\right)\,.
\nonumber
\end{eqnarray}

\section{4D Poincar\'e invariant solutions}
\label{sec:killing}

In this section, we discuss 4D Poincar\'e
invariant solutions of the $U(1)_R$-gauged orbifold SUGRA
presented in the previous section.
We first derive the Killing spinor equations and the energy
functional for generic 4D Poincar\'e invariant metric:
\begin{eqnarray}
ds^2=e^{2K(y)} \eta_{\underline\mu \underline\nu} (x)
dx^{\underline\mu} dx^{\underline\nu}-dy^2\,.
\label{eq:s:bkg}
\end{eqnarray}
and then consider some physical implications of the $U(1)_R$ FI terms
associated with a $U(1)_R$ charge $r\neq 0$.

\subsection{Killing conditions and 4D energy functional}

Applying the local SUSY transformations of the gravitinos,
gauginos, and the compensator and physical hyperinos~\cite{Fujita:2001bd},
we find the corresponding
Killing spinor conditions:
\begin{eqnarray}
\kappa &\equiv& \partial_yK-\frac{1}{3}M^I
\tilde{\cal Y}_I(i\sigma_3)^\dagger=0,
\nonumber \\
G^I &\equiv& \partial_y M^I -2 \left(
\tilde{\cal N}^{IJ}-\frac{1}{6}M^IM^J \right)
\tilde{\cal Y}_J=0,
\label{eq:orgggnkc} \\
F &\equiv& \partial_y\Phi-\Phi (V_y^N)^\dagger
-M^I(gt_I)\Phi(i\sigma_3)^\dagger
+\frac{1}{2}\Phi M^I\tilde{\cal Y}_I(i\sigma_3)^\dagger=0,
\nonumber \\
\tilde{F} &\equiv& \partial_y{\cal A}-{\cal A} (V_y^N)^\dagger
-\tilde{\cal F}\Big((i\sigma_3)^\dagger +\alpha^{-1}A^Z_y \Big)
\nonumber \\
&&-\,M^I(gt_I){\cal A}(i\sigma_3)^\dagger
+\frac{1}{2}{\cal A} M^I\tilde{\cal Y}_I(i\sigma_3)^\dagger=0\,,
\nonumber
\end{eqnarray}
where
\begin{eqnarray}
\tilde{\cal Y}_Z &=& -2\epsilon(y) \Big(
{\textstyle \frac{3}{2}}k{\cal A}^\dagger(i\sigma_3){\cal A}
+c \Phi^\dagger(i\sigma_3)\Phi \Big),
\nonumber \\
\tilde{\cal Y}_X &=& -2 \Big(
r{\cal A}^\dagger(i\sigma_3){\cal A}
+q \Phi^\dagger(i\sigma_3)\Phi \Big)
-\frac{1}{2}g\xi_{FI}\alpha^2(i\sigma_3)
(\delta(y)-\delta(y-\pi R))
\nonumber \\&&-2rM_{(4)}^2(i\sigma_3)(\Lambda_0 \delta(y)
+ \Lambda_\pi \delta(y-\pi R))\,.
\nonumber
\end{eqnarray}
and $M_{(4)}^2$ defined in Eq.~(\ref{eq:bigs}).
When all the above Killing conditions are satisfied,
there can be unbroken 4D $N=1$ supersymmetry with the
corresponding Killing spinor
\begin{eqnarray}
\epsilon_+(y) &=& \exp \left[
\frac{1}{2}(K(y)-K(0))\mathbf{1}_2+\int_0^y dz\, V_y^N(z)
\right] \epsilon_+(0)\,,
\nonumber
\end{eqnarray}
where $\epsilon_+=(\epsilon^{i=1}_R,\epsilon^{i=2}_L)$
and $\epsilon^i$ is the 5D supersymmetry transformation parameter.

The 4D energy density of a 4D Poincar\'e invariant configuration
is given by
\begin{eqnarray}
E &=& \int dy\,e^{4K}{\rm tr} \Bigg[
\frac{1}{4}a_{IJ}G^{I \dagger}G^J
-3|\kappa|^2+|F|^2-|\tilde{F}|^2
+|\Delta F|^2-|\Delta \tilde{F}|^2
\nonumber \\ && \qquad\qquad\qquad
-(F^\dagger \Delta F -\tilde{F}^\dagger \Delta \tilde{F} +\textrm{h.c.})
-2|V_{y\,{\rm sol}}|^2-\frac{1}{2}M^IV_{y\,{\rm sol}} [{\cal Y}_I,\,i\sigma_3]
+\Delta {\cal E} \Bigg]\,,
\nonumber
\end{eqnarray}
where
\begin{eqnarray}
\Delta F &=& -\Phi \Delta V_{y\,{\rm sol}}^\dagger\,, \qquad
\Delta \tilde{F} \ = \ -{\cal A} \Delta V_{y\,{\rm sol}}^\dagger
-\tilde{\cal F}\Big( (i\sigma_3)^\dagger +\alpha^{-1}A^Z_y \mathbf{1}_2 \Big)\,,
\nonumber \\
\Delta {\cal E} &=&
\frac{\Delta_D}{1+({\cal A}_+^{-2}-1) \Delta_D}
\Big( (V_{y\,{\rm sol}}^{(1)})^2 + (V_{y\,{\rm sol}}^{(2)})^2 \Big)
\mathbf{1}_2\,,
\nonumber
\end{eqnarray}
for $V_{y\,{\rm sol}}$, $\tilde{\cal F}$ and $\Delta V_{y\,{\rm sol}}$
given by Eqs.~(\ref{eq:osvy}), (\ref{eq:osaux}) and (\ref{eq:dv4sol}).
To arrive at this form  of 4D energy density starting from the 5D action
(\ref{eq:fullaction}), we have truncated the UV sensitive higher-order
boundary operators of ${\cal O}(\lambda \xi_{FI}^2)$,
${\cal O}(\lambda \Lambda_{0,\pi}^2)$
($\lambda=(\xi_{FI},\Lambda_{0,\pi},k,c)$)
whose precise value depend on how to regulate the $Z_2$-odd fields
at the boundaries.
The above form of the 4D energy density indicates that a field
configuration satisfying the Killing conditions:
\begin{eqnarray}
\kappa &=& G^I \ = \ F \ = \ \tilde{F} \ = \ 0\,,
\nonumber
\end{eqnarray}
as well as the stationary conditions:
\begin{eqnarray}
V_{y\,{\rm sol}} &=& 0\,, \qquad
[{\cal Y}_I,\,\sigma_3] \ = \ 0\,,
\nonumber
\end{eqnarray}
corresponds to a supersymmetric solution with vanishing vacuum energy.
A simple solution of the above stationary conditions is
\begin{eqnarray}
\Phi &=&
\left( \begin{array}{cc}
v(y) & 0 \\
0 & v(y)
\end{array} \right)\,,
\label{eq:stnrasz}
\end{eqnarray}
where $v$ is a real function of $y$.

If $\xi_{FI}=0$, the gaugino Killing conditions are simplified as
\begin{eqnarray}
G^I \equiv \frac{\partial M^I}{\partial \phi} D
= \frac{\partial M^I}{\partial \phi}\left[\partial_y\phi
+g^{\phi \phi}\frac{\partial M^I}{\partial \phi}
\tilde{\cal Y}_I(i\sigma_3)^\dagger\right] =0\,.
\nonumber
\end{eqnarray}
Since the physical implications of the FI coefficient $\xi_{FI}$ have
been studied extensively before~\cite{Ghilencea:2001bw,Barbieri:2002ic,
GrootNibbelink:2002wv,Correia:2004pz},
here we restrict ourselves to the case of $\xi_{FI}=0$ but
$r\neq 0$, while leaving the more general case with $\xi_{FI}\neq
0$ and $r\neq 0$ for future work. 
When $\xi_{FI}=0$, the Killing conditions and the form of the 4D energy
density can be simplified  under the ansatz
(\ref{eq:stnrasz}) for the physical hyperscalar field.
In this situation, we have
\begin{eqnarray}
M^I \tilde{\cal Y}_I(i\sigma_3)^\dagger
&=& {\cal P} \mathbf{1}_2\,, \qquad
\frac{\partial M^I}{\partial \phi}
\tilde{\cal Y}_I(i\sigma_3)^\dagger
\ = \ \partial_\phi( {\cal P} +\Xi) \mathbf{1}_2\,,
\nonumber
\end{eqnarray}
where
\begin{eqnarray}
{\cal P} &\equiv& -2 \left[\,  \left(
{\textstyle \frac{3}{2}}k\epsilon(y)\alpha +r \beta \right)
+v^2 \left( \big( {\textstyle \frac{3}{2}}k + c \big)\epsilon(y) \alpha
+\big( r+q \big) \beta \right) \,\right]\,,
\nonumber \\
\Xi &\equiv& -2rM_{(4)}^2 \beta
\Big(\,\Lambda_0\delta(y)+\Lambda_\pi\delta(y-\pi R)\,\Big)\,,
\nonumber
\end{eqnarray}
where we again ignore the UV-sensitive higher-dimensional
boundary operators. We then find the Killing parameters:
\begin{eqnarray}
\kappa &=& \partial_yK-\frac{1}{3} {\cal P}\,,
\nonumber \\
D &=& \partial_y\phi
+g^{\phi \phi}\partial_\phi({\cal P}+\Xi)\,,
\nonumber \\
F &=& \partial_yv-v \Big( q \beta + c \epsilon(y) \alpha
-\frac{1}{2}{\cal P} \Big)\,,
\nonumber \\
\tilde{F} &=&
\frac{v}{\sqrt{1+v^2}}F\,,
\nonumber
\end{eqnarray}
and also the 4D energy density:
\begin{eqnarray}
E &=& \int dy\,e^{4K}
\Bigg[ \frac{1}{2}g_{\phi \phi} D^2
+\frac{2}{1+v^2}F^2 -6\kappa^2 \Bigg]\,,
\nonumber
\end{eqnarray}
which tells us that the field configuration satisfying
\begin{eqnarray}
\kappa &=& D \ = \ F \ = \ 0\,,
\label{eq:kc}
\end{eqnarray}
corresponds to a supersymmetric vacuum solution of the theory.

\subsection{Vacuum solutions with $U(1)_R$ FI terms}
\label{sec:vacuum}
In this subsection, we discuss some aspects of the vacuum solution in
gauged $U(1)_R$ SUGRA on $S^1/Z_2$. Before going to the main analysis,
we briefly discuss the condition for the $U(1)_R$ anomaly cancellation.
To be complete, let us introduce boundary
chiral multiplets
$$
S_{0}^a=(z_0^a,\chi_{0}^a,f_{0}^a)\,, \quad
S_{\pi}^\alpha=(z_{\pi}^\alpha,\chi_{\pi}^\alpha,f_{\pi}^\alpha)\,,
$$
confined at $y=0$ and $y=\pi R$, respectively, and let $q_0^a$ and
$q_\pi^\alpha$ denote the $U(1)_X$ charge of their scalar components
$z_0^a$ and $z_\pi^\alpha$, respectively.
Here $\chi^a_0,\chi^\alpha_\pi$ and $f^a_0,f^\alpha_\pi$ denote the chiral
fermion and the complex auxiliary components of $S^a_0,S^\alpha_\pi$,
respectively.
Then the $U(1)_R^3$ and $U(1)_R$-gravity-gravity anomaly cancellation
conditions are given by
\begin{eqnarray}
&& (3+1)r^3+\sum_{\rm gaugino}r^3 + \sum_{\rm bulk} q^3
+\sum_a(q_0^a-r)^3
+\sum_\alpha(q_\pi^\alpha-r)^3
\ = \ 0\,,
\nonumber \\
&&(-21+1)r+\sum_{\rm gaugino}r + \sum_{\rm bulk} q
+\sum_a(q_0^a-r)
+\sum_\alpha(q_\pi^\alpha-r)
\ = \ 0\,,
\label{eq:amrcl}
\end{eqnarray}
where the first terms represent the contributions from the gravitino
and radino zero modes.

To see the effects of  $Z_2$-even $U(1)_R$-gauging, i.e., of $r\neq 0$,
let us first consider the simplest situation that
$\xi_{FI}=\Lambda_0=\Lambda_\pi =k=c=0$ and there is no $U(1)_X$-charged
boundary matter fields.
In this case, a unique solution of the Killing conditions (\ref{eq:kc})
is given by
$K=\phi=0$ and a {\it constant} hyperscalar VEV
$$
v=v_0 \equiv \pm \sqrt{-\frac{r}{r+q}}\,.
$$
Such hyperscalar VEV is allowed as long as
$q/r<-1$, which is in fact required in order for
the anomaly cancellation condition (\ref{eq:amrcl})  to be satisfied.

If one introduces boundary $U(1)_R$ FI terms
($\Lambda_{0,\pi}\neq 0$) into the above model, the  supersymmetric
vacuum solution is deformed as follows.
The corresponding Killing spinor conditions  are given by
\begin{eqnarray}
\partial_yK &=& -\frac{2\sqrt{2}}{3} \Big( r+(r+q)v^2 \Big)
\cosh^{2/3}(\phi) \tanh(\phi)\,,
\nonumber \\
\partial_y\phi &=& 2I(\phi)\left[\, r+(r+q)v^2+rM_{(4)}^2(v)
(\Lambda_0\delta(y)+\Lambda_\pi\delta(y-\pi R)) \,\right]\,,
\nonumber \\
\partial_yv &=& \sqrt{2}(r+q)(1+v^2)v \cosh^{2/3}(\phi) \tanh(\phi)\,,
\label{eq:nfigke}
\end{eqnarray}
where
\begin{eqnarray}
I(\phi)
=  \frac{\sqrt{2} \cosh^{2/3}(\phi)
\left( 2+ \cosh (2 \phi) \right)}{1+2\cosh (2\phi)}\,.
\nonumber
\end{eqnarray}
Obviously, the existence of
$\Lambda_0 \delta(y)+\Lambda_\pi \delta(y-\pi R)$
in the gaugino Killing condition enforces $\phi(y)$ to have a non-trivial
$y$-dependence, and thus $v(y)$ also.
In the limit that $|\,\phi(y)|\ll 1$ over the entire orbifold, which would
be the case if $|\,r|,|\,q|\ll 1$ (in the unit with $M_5=1$),
the gaugino and hyperino  Killing conditions can be approximated as
\begin{eqnarray}
\partial_y\phi &\simeq & 2\sqrt{2}\left[\, r+(r+q)v^2+rM_{(4)}^2(v)
(\Lambda_0\delta(y)+\Lambda_\pi\delta(y-\pi R)) \,\right]\,,
\nonumber \\
\partial_yv &\simeq & \sqrt{2}(r+q)v(1+v^2)\, \phi\,.
\label{ap-killing}
\end{eqnarray}
Then at leading order in $\delta v\equiv v-v_0$
($v_0=\pm \sqrt{-r/(r+q)}$)
which is presumed to be a small vacuum deformation, we find
\begin{eqnarray}
v&=& v_0+(Ae^{\omega y}+Be^{-\omega y})\,,
\\
\phi &=& -2\frac{|\,rv_0|}{rv_0}\sqrt{1+\frac{r}{q}}\,
\left( \, Ae^{\omega y}-Be^{-\omega y}\,\right)\,,
\nonumber
\end{eqnarray}
for $0<y<\pi R$, where $\omega=\sqrt{-8rq}$ \, and
\begin{eqnarray}
A&=& \frac{\sqrt{2}|\,rv_0|}{2v_0}\left(\frac{q}{r+q}\right)^{7/6}\left(
\frac{\Lambda_0+\Lambda_\pi e^{\omega \pi R}}{e^{2\omega \pi R}-1}\right)\,,
\nonumber \\
B&=& \frac{\sqrt{2}|\,rv_0|}{2v_0}\left(\frac{q}{r+q}\right)^{7/6}\left(
\frac{\Lambda_0+\Lambda_\pi e^{-\omega \pi R}}{1-e^{-2\omega \pi R}}\right)\,.
\nonumber
\end{eqnarray}
The above relations between the integration constants $A,B$ and the orbifold
radius $R$ show that once there exists a dynamics
to determine one (or both) of the boundary values of $v(y)$,
e.g., the boundary superpotentials of $\Phi$
which would determine $v(0)=v_0+A+B$ and/or
$v(\pi R)=v_0+Ae^{\omega \pi R}+Be^{-\omega \pi R}$,
one might be able to fix $R$
as well as to break $N=1$ SUSY through the combined effects of the
$U(1)_R$ FI terms and the boundary superpotentials.
Note that both $\phi(y)$ and $\delta v(y)$ are small over the entire orbifold
when $|\,r|,|q|\ll 1$, justifying the use of the approximate Killing conditions
(\ref{ap-killing}).

\begin{figure}[t]
\centerline{$I(\phi)$}
\begin{center}
\epsfig{figure=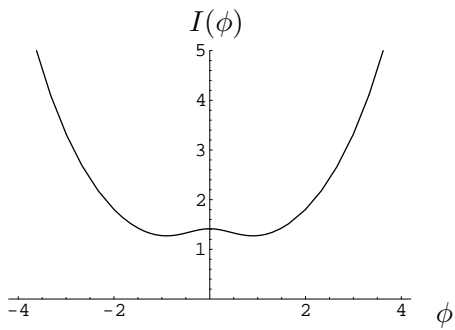,width=0.4\linewidth}$\phi$
\end{center}
\caption{Numerical value of $I(\phi)$ for $|\phi| \le 4$.}
\label{fig:1}
\end{figure}

As another example, we can consider the case that the $D$-flat
condition, i.e., the gaugino Killing condition, is satisfied through the
nonzero VEVs of the boundary chiral multiplets $S_0^a$ and/or $S_\pi^\alpha$,
while the bulk hyperscalars have vanishing VEVs.
For simplicity, let us again set
$\xi_{FI}=k=c=0$, while keeping $r$ and $\Lambda_{0,\pi}$ to be nonzero,
and assume the minimal form of the boundary K\"ahler potentials:
$K^{(0)}(S_0,\bar{S}_0)=\sum_a \bar{S}_0^aS_0^a$
and
$K^{(\pi)}(S_\pi,\bar{S}_\pi)=\sum_\alpha \bar{S}_\pi^\alpha S_\pi^\alpha$.
Then the gravitino and gaugino Killing conditions are given by
\begin{eqnarray}
\partial_y K&=&-\frac{2\sqrt{2}}{3}r\cosh^{2/3}(\phi)\tanh(\phi)\,,
\nonumber \\
\partial_y \phi &=& 2I(\phi)\left[\,
r+\lambda_0\delta(y)+\lambda_\pi\delta(y-\pi R)\,\right]\,,
\label{eq:dwg}
\end{eqnarray}
where $\lambda_0=(r+\sum_a q_0^a|z_0^a|^2)\Lambda_0$
and $\lambda_\pi=(r+\sum_\alpha q_\pi^\alpha|z_\pi^\alpha|^2) \Lambda_\pi$.
In the limit that $|\,\phi(y)|\ll 1$ over the entire orbifold,
this $D$-flat condition leads to
\begin{eqnarray}
\phi &\simeq & 2\sqrt{2}ry+\sqrt{2}\lambda_0,
\nonumber  \\
e^{2K} &\simeq &
\exp\left(\,-\frac{8}{3}r^2y^2-\frac{8}{3}\lambda_0 ry\,\right)\,,
\label{eq:sphiw0}
\end{eqnarray}
for $0<y<\pi R$.
Although derived under
the condition that $|\,\phi(y)|\ll 1$ for $0<y<\pi R$,
the above approximate solutions are valid
as long as $|\,\phi(y)|\lesssim 1$ for which
$I(\phi)$ is approximately a constant as can be seen from
Fig.~\ref{fig:1}.
(If $|\phi|\gtrsim 1$ near the boundary, the resulting boundary fluctuations
of $\beta(\phi)$ would be too large to be described by
orbifold field theory.)
In this case, the integrability condition $\oint dy \,\partial_y\phi=0$
determines  the  orbifold radius
(for given values of $\lambda_{0,\pi}$) as
$$
2\pi R \simeq  -\frac{\lambda_0+\lambda_\pi}{r}\,.
$$
Since $|\lambda_0+\lambda_\pi| \lesssim 1$ in order for the orbifold field
theory description to be valid, the above relation indicates that
$2 \pi R \lesssim 1/|r|$ for which the warp factor  $e^{2K}$ is
approximately a constant.
It also implies that once there exists a dynamics
to determine the VEVs of the boundary scalar fields $z^a_0$ and $z^\alpha_\pi$,
e.g., the boundary superpotentials,
one might be able to fix $R$ through the combined effects of the
$U(1)_R$ FI terms and the boundary superpotentials.

However the nontrivial profile of $\phi$ due to the $Z_2$-even $U(1)_R$
gauging can significantly alter the shape of the zero-mode wavefunction
$\Phi^{(0)}(y)$ of a $U(1)_X$-charged hypermultiplet.
To see this, let us consider a matter hypermultiplet with nonzero
hyperino $U(1)_X$ charge $q$ ($qr>0$) and vanishing VEV.
The corresponding zero mode obeys
$$
\partial_y \Phi^{(0)}(y) - m(y)\Phi^{(0)}(y) = 0\,,
$$
where
$$
m(y)=(3k/2+c) \epsilon(y) \alpha \big( \langle \phi(y) \rangle \big)
+(q+r) \beta \big( \langle \phi(y) \rangle \big)\,,
$$
in most general situation.
For $\xi_{FI}=k=c=0$ which leads to $\phi(y)$ given by (\ref{eq:sphiw0}),
one easily finds
\begin{eqnarray}
\Phi^{(0)}(y) &\simeq& \Phi^{(0)}(0)
\exp \left[ 2(q+r)(ry^2+\lambda_0 y) \right]\,.
\label{eq:gssn}
\end{eqnarray}

\begin{figure}[t]
\begin{center}
\begin{minipage}{0.35\linewidth}
   $\phi(y)$\\
   \centerline{\epsfig{figure=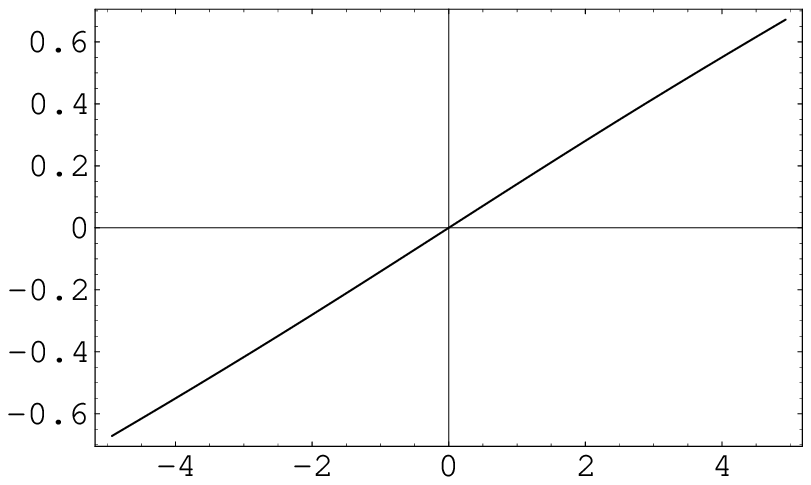,width=\linewidth}$y$}
\end{minipage}
\qquad
\begin{minipage}{0.35\linewidth}
   $\phi(y)$\\
   \centerline{\epsfig{figure=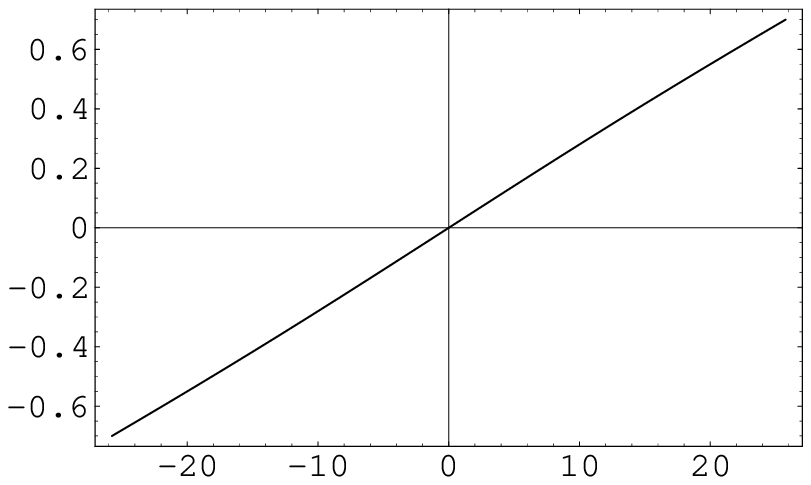,width=\linewidth}$y$}
\end{minipage}\\*[15pt]
\begin{minipage}{0.35\linewidth}
   $\log_{10} \Phi^{(0)}(y)/\Phi^{(0)}(\pi R)$\\
   \centerline{\epsfig{figure=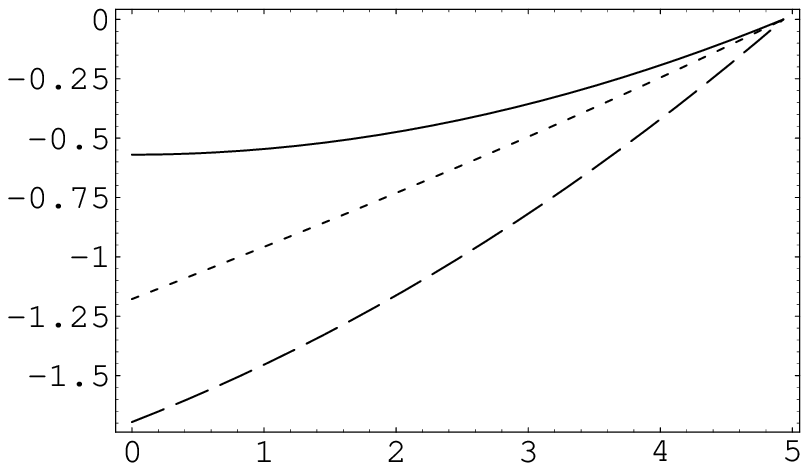,width=\linewidth}$y$}
\end{minipage}
\qquad
\begin{minipage}{0.35\linewidth}
   $\log_{10} \Phi^{(0)}(y)/\Phi^{(0)}(\pi R)$\\
   \centerline{\epsfig{figure=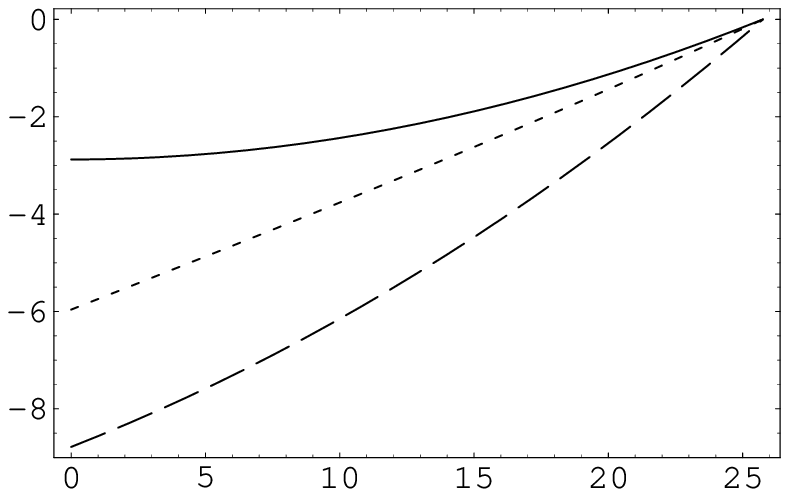,width=\linewidth}$y$}
\end{minipage}\\*[15pt]
\begin{minipage}{0.35\linewidth}
   \centerline{(a)\ $r=0.05$,\ $k=0$}
\end{minipage}
\qquad
\begin{minipage}{0.35\linewidth}
   \centerline{(b)\ $r=0.01$,\ $k=0$}
\end{minipage}
\end{center}
\caption{The profiles of  $\phi$ and the matter zero mode $\Phi^{(0)}$
for some cases with $k=0$ and $\lambda_0=0$.
Here we choose $\lambda_\pi=(r-1)/2$.
For the matter zero mode profile, the solid-,
dotted- and dashed-curves represent the case
with $(q,c)=(0.5,0)$, $(0,0.5)$ and $(0.5,0.5)$, respectively.
All parameters are given in the unit with $M_5=1$,
and all the curves are shown within $|y| \le \pi R$.}
\label{fig:2}
\end{figure}
For $\pi R|q(\lambda_0-\lambda_\pi)|\gtrsim 1$,
this matter zero mode would be quasi-localized
at one of the orbifold boundaries as
$$
\frac{\Phi^{(0)}(\pi R)}{\Phi^{(0)}(0)}\,\simeq\,
\exp \left[\,(q+r)\pi R (\lambda_0-\lambda_\pi)\,\right]\,.
$$
If $|q\pi R|$ is large enough, the resulting quasi-localization
of matter zero modes can generate small Yukawa couplings
in a natural manner as proposed in Ref.~\cite{Arkani-Hamed:1999dc}.
In Fig.~\ref{fig:2}, we show the profiles of $\phi(y)$ and the
corresponding hypermultiplet zero mode $\Phi^{(0)}$
in some cases with  $k=\lambda_0=0$
and $\lambda_\pi \ne 0$.
In this analysis, we consider also the case that the bare hypermultiplet
kink mass $c\epsilon(y)$ is non-vanishing.
Note that $\phi(y)$ takes a linear profile yielding the
Gaussian profile of the matter zero mode wavefunction.
The wave-function suppression at
$y=0$ caused by the gauged $U(1)_R$ FI terms can be as important
as the effect of the bare kink mass $c\epsilon(y)$.
As can be noticed from Fig.~\ref{fig:2}, one can have a stronger suppression
for smaller $r$ since then a larger orbifold radius is allowed.
We remark that if both $\lambda_0$ and $\lambda_\pi$ are non-vanishing,
the location of the minimum of the wave-function can be shifted.
The extreme case would be $\lambda_0=\lambda_\pi$ for which
the minimum is located at $y=\pi R/2$.

\begin{figure}[t]
\begin{center}
\begin{minipage}{0.35\linewidth}
   $\phi(y)$\\
   \centerline{\epsfig{figure=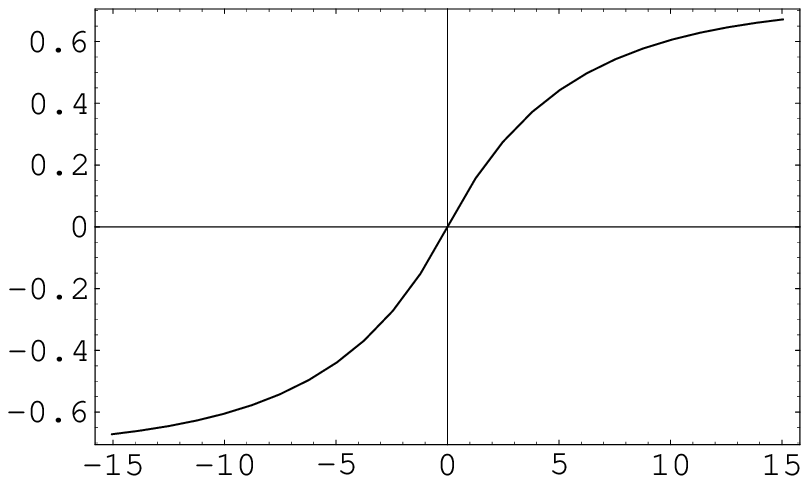,width=\linewidth}$y$}
\end{minipage}
\qquad
\begin{minipage}{0.35\linewidth}
   $\phi(y)$\\
   \centerline{\epsfig{figure=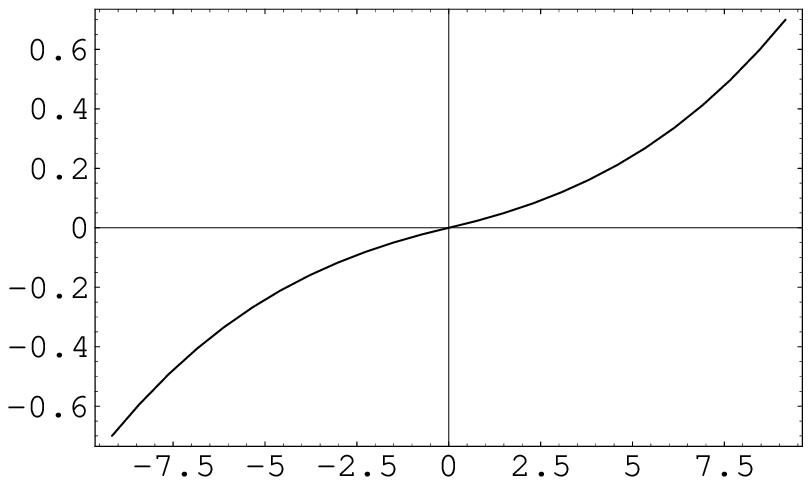,width=\linewidth}$y$}
\end{minipage}\\*[15pt]
\begin{minipage}{0.35\linewidth}
   $\log_{10} \Phi^{(0)}(y)/\Phi^{(0)}(\pi R)$\\
   \centerline{\epsfig{figure=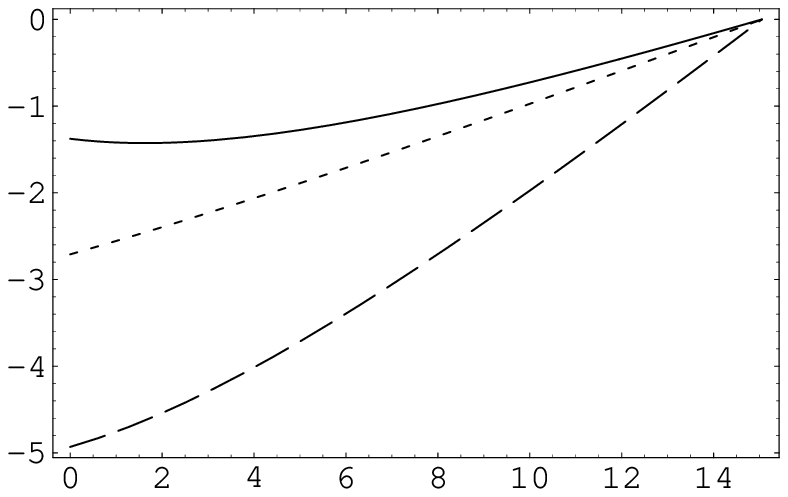,width=\linewidth}$y$}
\end{minipage}
\qquad
\begin{minipage}{0.35\linewidth}
   $\log_{10} \Phi^{(0)}(y)/\Phi^{(0)}(\pi R)$\\
   \centerline{\epsfig{figure=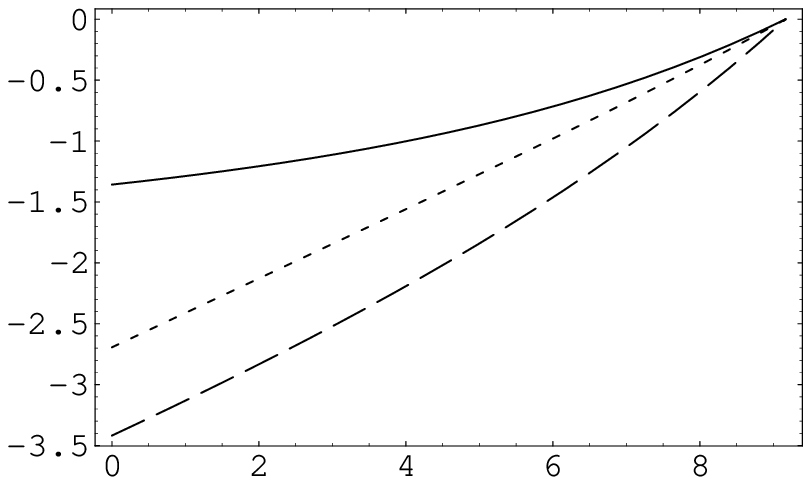,width=\linewidth}$y$}
\end{minipage}\\*[15pt]
\begin{minipage}{0.35\linewidth}
   \centerline{(a)\ $r=0.05$,\ $k=-0.1$}
\end{minipage}
\qquad
\begin{minipage}{0.35\linewidth}
   \centerline{(b)\ $r=0.01$,\ $k=0.1$}
\end{minipage}
\end{center}
\caption{The profiles of $\phi$ and $\Phi^{(0)}$
for $r,k \ne 0$, $\lambda_0=0$ and $\lambda_\pi=(r-1)/2$.
Again the solid-, dotted- and dashed-curves represent the case
$(q,c)=(0.5,0)$, $(0,0.5)$ and $(0.5,0.5)$, respectively.
Note that $K \simeq -ky$ in this supersymmetric solution.}
\label{fig:3}
\end{figure}

Fig.~\ref{fig:3} reveals the profiles of $\phi$ and $\Phi^{(0)}$ for
some cases with nonvanishing $k$.
Here $\phi$ has different profile for different sign of $k$,
Fig.~\ref{fig:3}(a) for $rk<0$ and (b) for $rk>0$.
Due to the effect of nonvanishing $k$, the orbifold
radius in Fig.~\ref{fig:3}(a) is larger than the
one in Fig.~\ref{fig:2}(a), while the radius in Fig.~\ref{fig:3}(b)
is smaller than the one in Fig.~\ref{fig:2}(b).
Consequently the wave-function suppression
becomes more (less) significant for $rk<0$ ($rk>0$)
compared to the case with $k=0$.

\section{Conclusion}

\label{sec:conclusion}
In this paper, we have studied  a 5D gauged $U(1)_R$ supergravity on
$S^1/Z_2$ in which both a $Z_2$-even $U(1)$ gauge field and the $Z_2$-odd
graviphoton take part in the $U(1)_R$ gauging.
Based on the off-shell 5D supergravity of Refs.~\cite{Fujita:2001bd},
we examined the structure of Fayet-Iliopoulos (FI) terms allowed by
such theory.
As expected, introducing a $Z_2$-even $U(1)_R$ gauging accompanies
new bulk and  boundary FI terms in addition to
the known integrable boundary FI term which could be present
in the absence of any gauged $U(1)_R$ symmetry.
The new (non-integrable) boundary FI terms  originate from the $N=1$
boundary supergravity, and thus are free from the bulk supergravity
structure in contrast to the integrable boundary FI term which is
determined by the bulk structure of 5D
supergravity~\cite{Ghilencea:2001bw,Barbieri:2002ic,
GrootNibbelink:2002wv,Correia:2004pz}.

We have examined some physical consequences of the $Z_2$-even $U(1)_R$
gauging in several simple cases.
It is noted that the FI terms of gauged $Z_2$-even $U(1)_R$  can lead
to an interesting deformation of vacuum structure which can affect
the quasi-localization of the matter zero modes in extra dimension
and also the SUSY breaking and radion stabilization.
Thus the 5D gauged $U(1)_R$ supergravity on orbifold has a rich
theoretical structure which may be useful for understanding some
problems in particle physics such as the Yukawa
hierarchy~\cite{Arkani-Hamed:1999dc}
and/or the supersymmetry breaking.
When one tries to construct a realistic particle physics model within
gauged $U(1)_R$ supergravity, the most severe constraint comes from the
anomaly cancellation condition~\cite{Barbieri:1982ac,Kitazawa:1999su}.
In some cases the Green-Schwarz mechanism might
be necessary to cancel the anomaly,  which may introduce another
type of FI term into the theory~\cite{Dudas:2004ni}.
These issues will be studied in future works.

\subsection*{Acknowledgement}
We would like to thank Ian-Woo~Kim and
Keisuke~Ohashi for useful discussions and comments.
This work is supported by KRF PBRG 2002-070-C00022
(HA,KC) and the Center for High Energy Physics of
Kyungbook National University (KC).


\begin{thebibliography}{99}

\bibitem{O'Raifeartaigh:1975pr}
L.~O'Raifeartaigh,
Nucl.\ Phys.\ B {\bf 96}, 331 (1975).

\bibitem{Fayet:1974jb}
P.~Fayet and J.~Iliopoulos,
Phys.\ Lett.\ B {\bf 51}, 461 (1974).

\bibitem{Barbieri:1982ac}
R.~Barbieri, S.~Ferrara, D.~V.~Nanopoulos and K.~S.~Stelle,
Phys.\ Lett.\ B {\bf 113}, 219 (1982);
%
A.~H.~Chamseddine and H.~K.~Dreiner,
Nucl.\ Phys.\ B {\bf 458}, 65 (1996)
[hep-ph/9504337];
%
D.~J.~Castano, D.~Z.~Freedman and C.~Manuel,
Nucl.\ Phys.\ B {\bf 461}, 50 (1996)
[hep-ph/9507397];
%
P.~Binetruy, G.~Dvali, R.~Kallosh and A.~Van Proeyen,
Class.\ Quant.\ Grav.\  {\bf 21}, 3137 (2004)
[hep-th/0402046].

\bibitem{Green:1984sg}
M.~B.~Green and J.~H.~Schwarz,
Phys.\ Lett.\ B {\bf 149}, 117 (1984).

\bibitem{Altendorfer:2000rr}
R.~Altendorfer, J.~Bagger and D.~Nemeschansky,
Phys.\ Rev.\ D {\bf 63}, 125025 (2001)
[hep-th/0003117];
%
T.~Gherghetta and A.~Pomarol,
Nucl.\ Phys.\ B {\bf 586}, 141 (2000)
[hep-ph/0003129];
%
A.~Falkowski, Z.~Lalak and S.~Pokorski,
Phys.\ Lett.\ B {\bf 491}, 172 (2000)
[hep-th/0004093];
%
K.~Choi, H.~D.~Kim and I.~W.~Kim,
JHEP {\bf 0211}, 033 (2002)
[hep-ph/0202257];
%
K.~Choi, H.~D.~Kim and I.~W.~Kim,
JHEP {\bf 0303}, 034 (2003)
[hep-ph/0207013].

\bibitem{Randall:1999ee}
L.~Randall and R.~Sundrum,
Phys.\ Rev.\ Lett.\  {\bf 83}, 3370 (1999)
[hep-ph/9905221].

\bibitem{Fujita:2001bd}
T.~Fujita, T.~Kugo and K.~Ohashi,
Prog.\ Theor.\ Phys.\  {\bf 106}, 671 (2001)
[hep-th/0106051];
%
T.~Kugo and K.~Ohashi,
Prog.\ Theor.\ Phys.\  {\bf 108}, 203 (2002)
[hep-th/0203276].

\bibitem{Ghilencea:2001bw}
D.~M.~Ghilencea, S.~Groot Nibbelink and H.~P.~Nilles,
Nucl.\ Phys.\ B {\bf 619}, 385 (2001)
[hep-th/0108184].

\bibitem{Barbieri:2002ic}
R.~Barbieri, R.~Contino, P.~Creminelli, R.~Rattazzi and C.~A.~Scrucca,
Phys.\ Rev.\ D {\bf 66}, 024025 (2002)
[hep-th/0203039].

\bibitem{GrootNibbelink:2002wv}
S.~Groot Nibbelink, H.~P.~Nilles and M.~Olechowski,
Phys.\ Lett.\ B {\bf 536}, 270 (2002)
[hep-th/0203055]; 
%
S.~Groot Nibbelink, H.~P.~Nilles and M.~Olechowski,
Nucl.\ Phys.\ B {\bf 640}, 171 (2002)
[hep-th/0205012];
%
D.~Marti and A.~Pomarol,
Phys.\ Rev.\ D {\bf 66}, 125005 (2002)
[hep-ph/0205034];
%
H.~Abe, T.~Higaki and T.~Kobayashi,
Prog.\ Theor.\ Phys.\  {\bf 109}, 809 (2003)
[hep-th/0210025];
%
T.~Hirayama and K.~Yoshioka,
JHEP {\bf 0401}, 032 (2004)
[hep-th/0311233];
%
T.~Kobayashi and K.~Yoshioka,
JHEP {\bf 0411}, 024 (2004)
[hep-ph/0409355].


\bibitem{Correia:2004pz}
F.~P.~Correia, M.~G.~Schmidt and Z.~Tavartkiladze,
hep-th/0410281.

\bibitem{Arkani-Hamed:1999dc}
N.~Arkani-Hamed and M.~Schmaltz,
Phys.\ Rev.\ D {\bf 61}, 033005 (2000)
[hep-ph/9903417];
%
E.~A.~Mirabelli and M.~Schmaltz,
Phys.\ Rev.\ D {\bf 61}, 113011 (2000)
[hep-ph/9912265];
%
D.~E.~Kaplan and T.~M.~P.~Tait,
JHEP {\bf 0111}, 051 (2001)
[hep-ph/0110126];
%
M.~Kakizaki and M.~Yamaguchi,
Int.\ J.\ Mod.\ Phys.\ A {\bf 19}, 1715 (2004)
[hep-ph/0110266];
%
N.~Haba and N.~Maru,
Phys.\ Rev.\ D {\bf 66}, 055005 (2002)
[hep-ph/0204069];
%
A.~Hebecker and J.~March-Russell,
Phys.\ Lett.\ B {\bf 541}, 338 (2002)
[hep-ph/0205143];
%
K.~Choi, D.~Y.~Kim, I.~W.~Kim and T.~Kobayashi,
Eur.\ Phys.\ J.\ C {\bf 35}, 267 (2004)
[hep-ph/0305024];
%
K.~Choi, I.~W.~Kim and W.~Y.~Song,
Nucl.\ Phys.\ B {\bf 687}, 101 (2004)
[hep-ph/0307365];
%
H.~Abe, K.~Choi, K.~S.~Jeong and K.~i.~Okumura,
JHEP {\bf 0409}, 015 (2004)
[hep-ph/0407005].

\bibitem{Bergshoeff:2000zn}
E.~Bergshoeff, R.~Kallosh and A.~Van Proeyen,
JHEP {\bf 0010}, 033 (2000)
[hep-th/0007044].

\bibitem{Kugo:2000af}
T.~Kugo and K.~Ohashi,
Prog.\ Theor.\ Phys.\  {\bf 105}, 323 (2001)
[hep-ph/0010288].

\bibitem{PaccettiCorreia:2004ri}
F.~Paccetti Correia, M.~G.~Schmidt and Z.~Tavartkiladze,
hep-th/0408138;
%
H.~Abe and Y.~Sakamura,
JHEP {\bf 0410}, 013 (2004)
[hep-th/0408224].

\bibitem{Kitazawa:1999su}
N.~Kitazawa, N.~Maru and N.~Okada,
Phys.\ Rev.\ D {\bf 62}, 077701 (2000)
[hep-ph/9911251];
%
N.~Kitazawa, N.~Maru and N.~Okada,
Nucl.\ Phys.\ B {\bf 586}, 261 (2000)
[hep-ph/0003240].

\bibitem{Dudas:2004ni}
E.~Dudas, T.~Gherghetta and S.~Groot Nibbelink,
Phys.\ Rev.\ D {\bf 70}, 086012 (2004)
[hep-th/0404094].


\end{thebibliography}
\end{document}